\documentclass[conference]{IEEEtran}
\IEEEoverridecommandlockouts
\usepackage{cite}
\usepackage{amsmath,amssymb,amsfonts}
\usepackage{algorithmic}
\usepackage{graphicx}
\usepackage{textcomp}
\usepackage{xcolor}

\usepackage{booktabs}
\usepackage{multirow}
\usepackage{color, colortbl}
\usepackage{xcolor}
\definecolor{Gray}{gray}{0.92}

\def\BibTeX{{\rm B\kern-.05em{\sc i\kern-.025em b}\kern-.08em
    T\kern-.1667em\lower.7ex\hbox{E}\kern-.125emX}}
\begin{document}

\title{Near Optimal Per-Clip Lagrangian Multiplier Prediction in HEVC}

\author{\IEEEauthorblockN{Daniel J Ringis.  Fran\c{c}ois Piti\'e and Anil Kokaram}
\IEEEauthorblockA{\textit{Sigmedia Group, Electronic and Electrical Engineering Dept.} \\
\textit{Trinity College Dublin}\\
Dublin, Ireland \\
{\{ringisd, pitief, anil.kokaram\} @tcd.ie} }
}

\maketitle

\begin{abstract}
The majority of internet traffic is video content. This drives the demand for video compression  to deliver high quality video at low target bitrates. Optimising  the parameters of a video codec for a specific video clip (per-clip optimisation) has been shown to yield significant bitrate savings. In previous work we have shown that per-clip optimisation of the Lagrangian multiplier leads to up to 24\% BD-Rate improvement. A key component of these algorithms is modeling the R-D characteristic across the appropriate bitrate range. This is computationally heavy as it usually involves repeated video encodes of the high resolution material at different parameter settings. This work focuses on reducing this computational load by deploying a NN operating on lower bandwidth features. Our system achieves BD-Rate improvement in approximately 90\% of a large corpus with comparable results to previous work in direct optimisation.
\end{abstract}

\begin{IEEEkeywords}
Video Compression, Video Codecs, Adaptive Encoding
\end{IEEEkeywords}

\section{Introduction}
With video content comprising over 80\% of all internet traffic~\cite{cisco}, there is an ever growing need for video compression for user generated content (UGC). The recent impact of the pandemic emphasises the importance of developing new techniques. The modern compression schemes H.265/6 (HEVC)~\cite{sullivan2012overview, zhang2019overview, bross2020versatile} and VP9 and AV1 \cite{mukherjee2013latest, chen2018overview}, have all been thrown into the spotlight.

Striking a balance between the rate and distortion of a video clip is a key challenge in video coding. Since 1998~\cite{sullivan1998rate}, it has been accepted practice to represent this trade-off in terms of a cost $J$ as follows:
\begin{equation}
    J = D + \lambda R \label{rd}
\end{equation}
Here, a distortion, $D$ and a rate $R$ are combined through the action of a Lagrangian multiplier, $\lambda$. The Lagrangian multiplier controls emphasis on either rate minimisation or quality improvement. This has an impact throughout the codec, as the idea is applied to many internal operations  e.g.  motion vectors, block type (Skipped/Intra/Inter), and bit allocation at the frame and clip levels~\cite{wiegand2001lagrange}.

The choice of \(\lambda\) was determined experimentally in the early days of codec research \cite{ortega1998rate}. This parameter can be considered as a kind of hyperparameter in a hybrid codec as it affects so many other decisions. This was originally implemented for the reference implementation of the H.263 standard and then modified for the implementations of H.264 and H.265 (HEVC)~\cite{hevcOverview}. The approach was to select a value based on a cohort of examples such that the performance was optimised {\em on average} across the set. That set was quite small, less than 5 examples. While this small dataset may have been an adequate representation of videos when developed, it may not be ideal with the large range of user generated video content today.

It has been recognised that choosing coding parameters which are optimised for a particular video lead to bitrate savings~\cite{katsavounidis2018video,covell2016optimizing}. This is because the statistics of video clips varies greatly over any corpus. In our previous work~\cite{EIRingis, SPIERingis}, we have shown that direct optimisation methods can lead to bitrate gains per clip of up to 20\%. Unfortunately, this bitrate savings came at a high computational cost, i.e. in excess of fifty video encodes per clip. At scale, this cost is prohibitive.

This work therefore makes two principal contributions i) development of a more computationally efficient optimiser for the Lagrangian based on Neural Networks and ii) exploration of the gains achievable on a per clip basis.

\section{Previous Work on RD Optimisation}

The rate distortion algorithm \cite{wiegand1996rate} establishes a balance between the quality of the media and the transmission or storage capacities of the medium. As mentioned above, seminal work of Sullivan and Wiegand~\cite{sullivan1998rate} laid the foundation for an experimental approach to choosing an appropriate \(\lambda\). This work established a relationship between the quantisation step size \(Q\) and the distortion \(D\) in a frame. Through minimising $J$, this leads to  a relationship between \(\lambda\) and \(Q\) expressed as $\lambda = 0.85 \times Q^2$. Updates to those experiments\cite{hevcOverview} then yielded similar relationships in the most common implementations of H.264 and H.265 (HEVC).  Introduction of bi-directional (B) frames led to different  constants and three different relationships were established for each of the Intra (I), $\lambda_{I} =( 0.57 )2^{(Q-12)/3}$, Predicted (P), $\lambda_{P} = (0.85) 2^{(Q-12)/3}$, and B frames, $\lambda_{B}  = ( 0.68 ) \max(2, \min(4, (Q - 12)/6))  2^{(Q-12)/3}$.

\subsection{Per Clip Optimisation}

There exists a limited amount of work on adaptation of $\lambda$ in the rate distortion equation. In all cases, results support the idea that adjusting  \(\lambda\) leads to improvement in codec performance {\em per clip}. The idea is typically to adjust $\lambda$ away from the codec default by using a constant $k$ as follows.
\begin{equation}
   \lambda_{\textrm{new}} = k \times \lambda_{\textrm{orig}}\label{kfactor}
\end{equation}
where $\lambda_{\textrm{orig}}$ is the default Lagrangian multiplier estimated in the video codec, and  $\lambda_{\textrm{new}}$ is the updated Lagrangian.

Ma et al~\cite{ma2016adaptive} used a Support Vector Machine to determine \(k\). The {\em perceptual} feature set included scalars representing Spatial Information and Temporal Information, and a texture feature exploiting a Gray Level Concurrence Matrix. Details can be found in \cite{ma2016adaptive}. Their focus was on {\em Dynamic textures} and they used the DynTex dataset of 37 sequences. They reported up to 2dB improvement in PSNR and 0.05 improvement in SSIM at equal bitrates. Hamza et al~\cite{hamza2019parameter} also take a classification approach but using gross scene classification into indoor/outdoor/urban/non-urban classes. They then used the same $k$ for each class. 
Their work used the Derfs dataset and reported up to 6\% BD-Rate improvement.

Zhang and Bull~\cite{zhang_bull} used a single feature ${D}_{P}/{D}_{B}$, the ratio between the MSE of P and B frames. This feature gives some idea of temporal complexity. Experiments based on the DynTex database yielded up to 7\% improvement in BD-Rate. They modified $\lambda$ implicitly by adjusting the quantiser parameter $Q$. Papadopoulos et al~\cite{Papadopoulos} exploited this and applied an offset to Q, in HEVC, based on the ratio of the distortion in the P and B frames. Each QP was updated from the previous Group of Pictures (GOP) using $ \textrm{QP} = a \times ({D}_{P}/{D}_{B}) - b $
where $a,b$ are constants determined experimentally.  This lead to an average BD-Rate improvement of 1.07\% on the DynTex dataset, with up to 3\% BD-Rate improvement achieved for a single sequence.

Yang et al~\cite{yang2017perceptual} used a combination of features instead of just the MSE ratio above. In their work, they used a perceptual content measurement $S$ to model $k$ with a straight line fit $ k = aS - b$.
Here again $a,b$ were determined experimentally using a corpus of the Derfs dataset. The loss in complexity of the fit is compensated for by the increase in complexity of the feature. They report a BD-Rate improvement of up to 6.2\%.
Recently, John et al\cite{john2020rate} proposed a machine learning method to classify Rate Distortion characteristics of a clip in order to select the correct operating point for that clip. In this work, they cluster videos based on their RD-Curve operating points and determines the appropriate encoding parameters for each video based on the model developed. 

\subsection{Direct Optimisation of $k$\label{dopt}}

\begin{figure}
    \centering
    \includegraphics[width=0.49\linewidth]{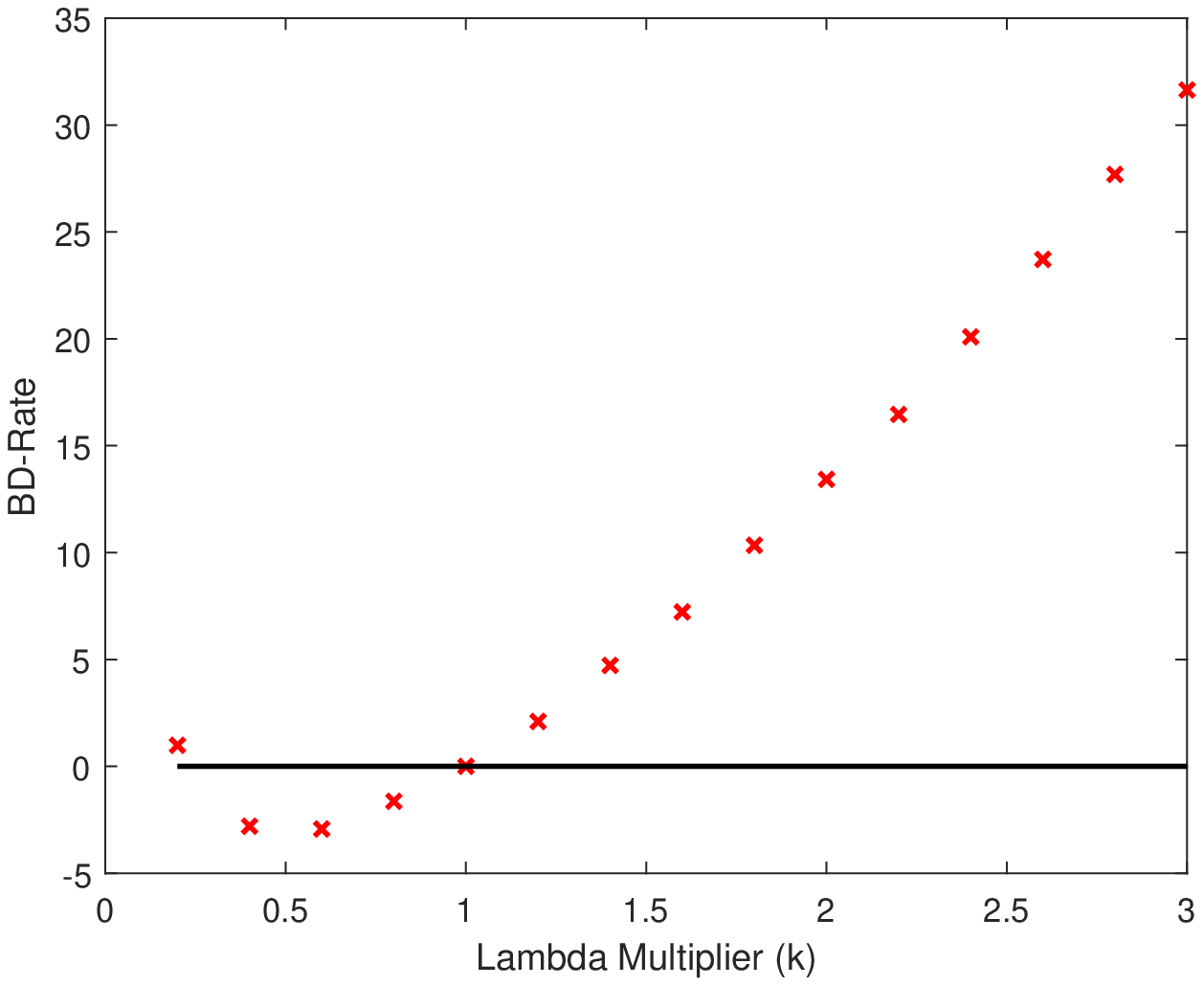}
    \includegraphics[width=0.49\linewidth]{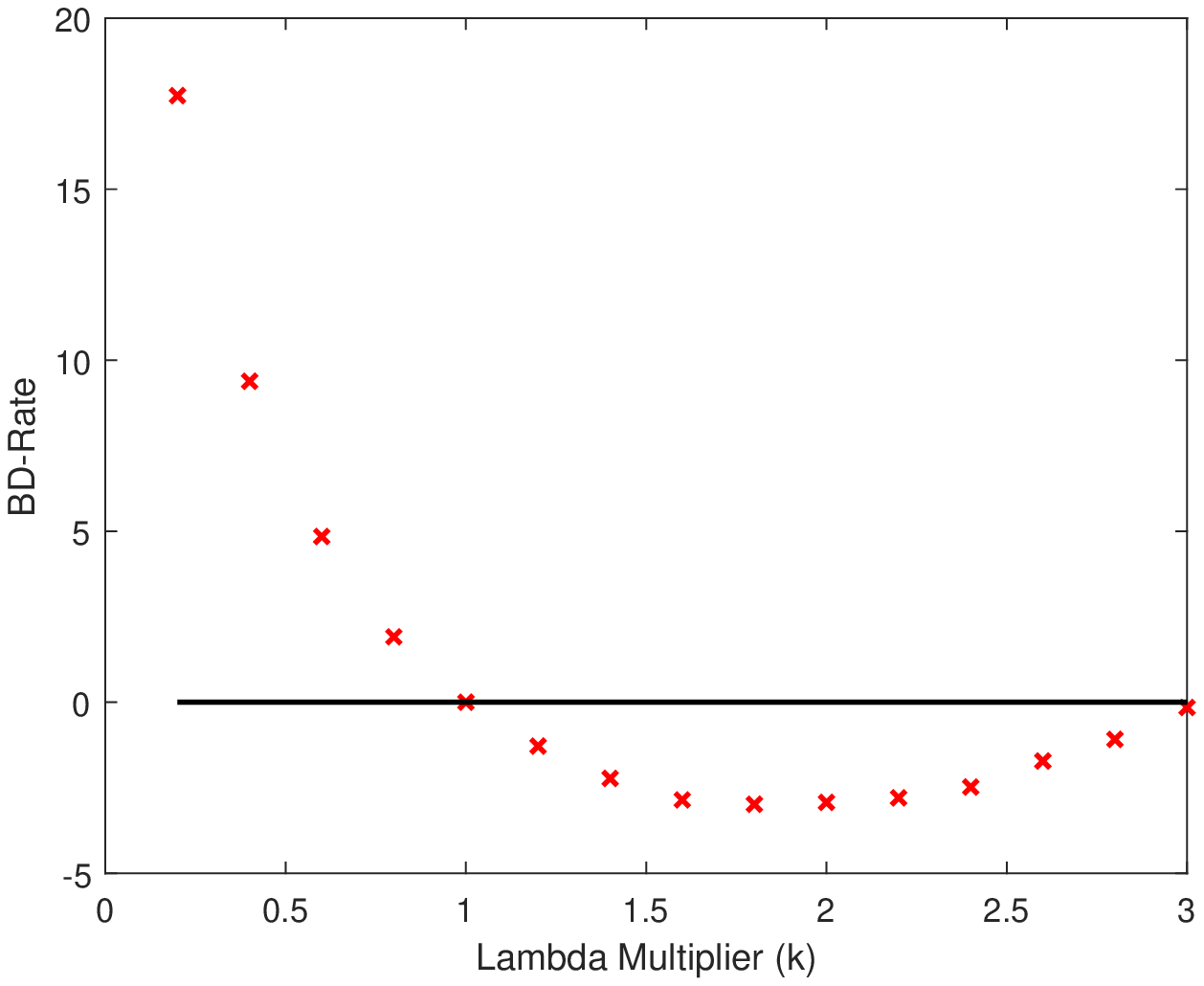}
    \caption{BD-Rate vs $k = [0.2:0.2:3]$ for two different clips
      (LiveMusic\_1080P-6d1a and NewsClip\_720P-7e56). The best performing BD-Rate gain is achieved at a
      different value for $k$ in each clip. Hence per-clip optimisation is
      sensible. The curve shape shows a global minimum implying that classic
      optimisation strategies would be successful.}
    \label{exampleBDk}
\end{figure}

As shown in Equation \ref{kfactor}, previous work attempts to adjust the rate control parameter $\lambda$ adaptively in some way. In order to show that there is something to be gained in varying $k$ per clip, Figure \ref{exampleBDk} shows BD-Rate vs $k$ for two different clips. There are two observations to be made. Firstly, the relationship between $k$ and BD-Rate varies substantially between clips. This reinforces the view that the traditional one size fits all approach is sub-optimal. Secondly, the general shape of the curves shows a global minimum. That minimum represents the maximum BD-Rate improvement available.

In recently published works \cite{EIRingis, SPIERingis} we introduce the use of direct optimisation to maximise BD-R and evaluated it on an expanded dataset, primarily based on the  YouTube-UGC dataset \cite{wang2019youtube}. The dataset was complemented with clips from other publicly available datasets, including the Netflix dataset (Chimera and El Fuente)\cite{netflixdb}, DynTex dataset\cite{dyntex}, MCL\cite{MCL} and Derfs dataset\cite{derf}. Multiple DASH segments (clips) of 5 seconds (150 frames) were created from each sequence. The dataset comprises a total of 9,746 video clips at varying resolutions with a wide range of video content, representative of typical usage. 

\begin{figure}
    \centering
    \includegraphics[height=0.7 \columnwidth]{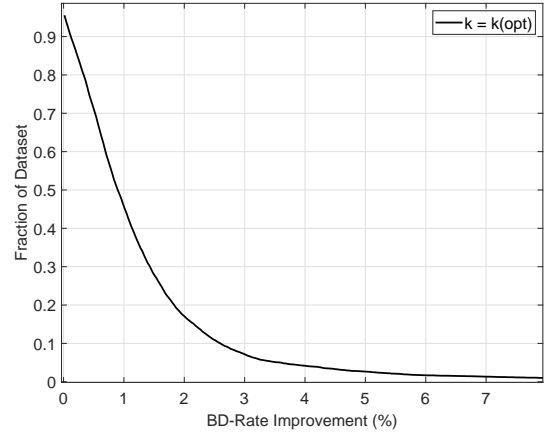}
    \caption{BD-Rate improvement vs fraction of dataset which achieves that improvement or better. We can see from this curve that the direct optimisation method was capable of achieving BD-Rate improvement in 95\% of the clips in the corpus. We can also read off of this graph what fraction of the clips had a 1\% or better BD-Rate improvement (47\%) and what fraction of the clips had a 5\% or better improvement (3\%). }
    \label{DirectOpt}
\end{figure}

In Figure \ref{DirectOpt}, we report the BD-Rate improvement found in \cite{EIRingis} to be achievable by such a direct optimisation of the Lagrangian Multiplier. In this graph, we see the BD-Rate improvement on the x-axis and the fraction of the dataset which was able to achieve that BD-Rate improvement or better. Ideally, we would want this curve to be as close to the top right as possible, as that would indicate all clips achieved high BD-Rate improvements. We will be using these results as our ground truth and upper bound for the deep learning systems described later in this paper. The key result of this work is that 95\% of the clips had some BD-Rate improvement over the default Lagrangian Multiplier, with 46\% of them showing a BD-Rate improvement of 1\% or better. 

\subsection{Machine-Learning Prediction of $k$\label{mlopt}}

The direct optimisation method \cite{EIRingis} outlined in the previous section, provides positive results, however, it is computationally expensive. It takes on average 11.7 iterations, and, as each BD-rate computation takes 5 video encodes, finding the optimal $k$ takes $11.7 \times 5 \approx 60$ video encodes. This is manageable at small resolutions, but is computationally intensive at modern resolutions (720p and higher) and {\em at scale}. 

Because of this, we have proposed in \cite{SPIERingis} to reduce the computational cost by using proxies (low resolutions, older codecs) to calculate $k$ in practice. We were able to get a reduction of $3.1\times$, while achieving 83\% of the gain found by direct optimisation. To attempt at further reducing the computational cost, we also trained a machine learning system (Random Forest) based on encoding information given by the x265 encoder at CRF 33. The results, reported in the second row of Table \ref{MLSummary}, show that the Random Forest model leads to an average BD-Rate improvement of 0.19\% and is still some way away from the potential 1.87\% improvement given by direct optimisation. 

\section{Is the Current Lagrangian Multiplier Optimal Across a Modern Corpus? }

\begin{figure}
    \centering
    \includegraphics[height=0.7 \columnwidth]{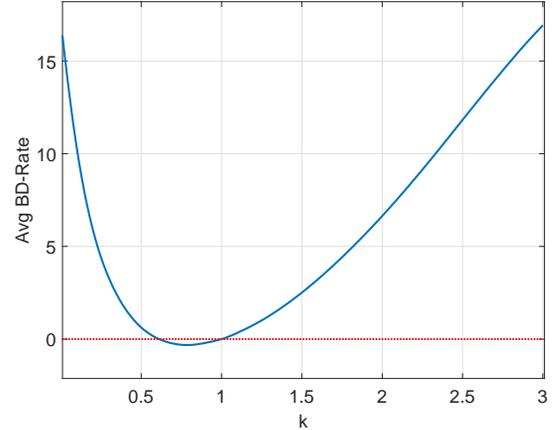}
    \caption{Average BD-Rate for a given k. Values of $k$ in the range $[0.7,0.8]$ give a clear average improvement (negative BD-rate) across the entire dataset. }
    \label{KAvg}
\end{figure}

\begin{figure}
    \centering
    \includegraphics[height=0.7 \columnwidth]{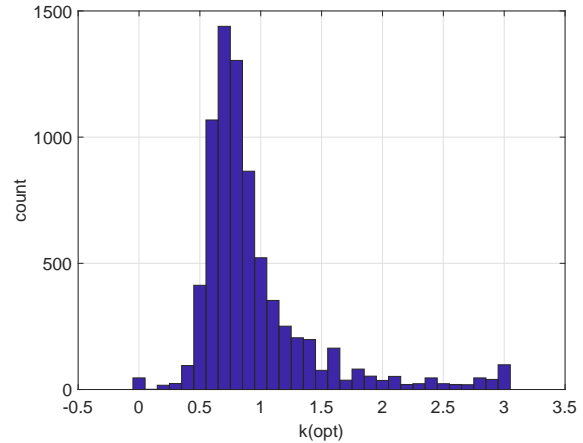}
    \caption{Histogram of optimum $k$ (and by extension optimum Lagrangian multiplier). The optimum is most frequently found in the $k$=0.7-0.8 range.}
    \label{kHist}
\end{figure}

Before exploring how to predict a better per-clip $k$, we propose first to investigate if $k=1$, as set in the MPEG standard, is indeed optimal for our corpus. What we propose here is to exhaustively calculate the BD-Rate over our corpus for values of $k$ in the range $k$=0:0.001:3. The average BD-Rate improvement for a given $k$ can be seen in Figure \ref{KAvg}. The figure shows an improvement where $k$ is approximately 0.7:0.8, with a best average improvement coming at $k=0.782$. This also matches with the histogram of the optimal $k$ values found with direct optimisation on the corpus videos (see Figure \ref{kHist}). The following command invocation was used for all video encodes:

\texttt{x265 --input SEQ.y4m --crf <XX> --tune-PSNR --psnr --output OUT.mp4 }

where {\tt SEQ} and  {\tt OUT}  are the filenames for the raw input file and output encoded video. Clips were encoded with {\tt <XX>}  in the range 22:5:42.

 For $k=0.782$, about two-thirds of the clips have an improvement but it is worse for one-third of them (see Table \ref{MLSummary}). Overall, we get an average gain of 0.32\% across our corpus. This indicates that the Lagrangian Multiplier for HEVC may not be the best. This would imply that for HEVC, we should adjust the Lagrangian Multiplier to be $0.782 \times$ its current value. However, this may be specific to our corpus and the operating points which we are using.

\begin{figure}
    \centering
    \includegraphics[height=0.7 \columnwidth]{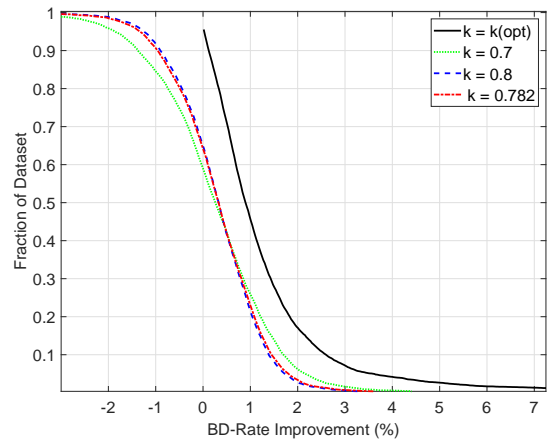}
    \caption{CDF plots of the BD-Rate improvement across our corpus when using $k$ = 0.7, 0.8 or 0.782. Ideally, we would want these plots to be as close as possible to the top right and coincide with the optimal per-clip plot (in black). We see for $k$ = 0.7, 0.8 or 0.782 positive BD-Rate improvements for 60-70\% of the clips but worse performance for 30-40\%.  }
    \label{SanityCheck}
\end{figure}

\section{Deep Learning for per clip Lagrangian Multiplier optimisation}

We propose in this section to further explore the machine learning approach proposed in our previous work \cite{SPIERingis}. The main contributions are 1) to extend the feature set to per frame encoding statistics instead of global sequence statistics, 2) to include semantic visual features, and lastly 3) to employ a Deep Neural Network (DNN) instead of Random Forest. 

\subsection{Input Features}

For this work, we collect two sets of features from each video from our corpus. As with our previous work \cite{SPIERingis}, our first set of features comes from the statistics generated in the x265 encoder report. 
\begin{table}[t]

\centering
\caption{\label{featureTable}List of Encoding Features used in Deep Learning system}
\begin{tabular}{|l|l|l|}
\hline
\multicolumn{1}{|c|}{\textbf{Features Per Clip}} & \multicolumn{1}{c|}{\textbf{Features Per Frame}} \\ \hline
Average Bitrate                                  & Bits                                             \\ \hline
Average PSNR (Y, U, V, Global)                   & I/P cost Ratio                                   \\ \hline
Average I, P and B frame count                   & PSNR (Y,U,V)                                     \\ \hline
Bitrate for I, P and B frames                    & Avg Chroma Distortion                            \\ \hline
I Frame PSNR (Y,U,V)                             & Average Residual Energy                          \\ \hline
P Frame PSNR (Y,U,V)                             & Avg Luma Level                                   \\ \hline
B Frame PSNR (Y,U,V)                             & Avg Cb Level                                     \\ \hline
                                                 & Avg Cr Level                                     \\ \hline
\end{tabular}
\end{table}
We expand from \cite{SPIERingis} by also including the per frame features (see full list in Table \ref{featureTable}). This greatly expands the feature vector from 47 to 1523 features. The features are collected for a clip encoded with no modifications to the codec, at CRF 33. 
 
In addition to these features, we also use the VGG16 network to generate semantic visual features about a single frame from each video sequence. We observed, indeed, that the nature of the clips content influenced the Lagrangian optimisation. The 1000 features from the 3c7 layer of VGG16 are used on a downsampled ($224 \times 224$) median (75th) frame of the sequence. These two feature vectors are then concatenated into a combined feature vector of size 2523 for our model. Using the VGG features and NN architecture does come at a cost of increased complexity, but this may be preferable to multiple video encodes required for the direct optimisation method.
 
\begin{table}[t]
\centering
\caption{\label{MLSummary} Summary of {\bf BD-R Gains} results. This shows the $\%$ of clips which have any improvement, $\> 0.1\%$ and $\>1\%$ improvement, as well as the best and average positive BD Rate improvement  }
\setlength{\tabcolsep}{4pt}
\begin{tabular}{m{9em}rrrrr}
\toprule
\multirow{2}{*}{\textbf{Method}} & \multicolumn{3}{c}{\bf Clips with BD-R Gain of} & {\textbf{Best}} & {\textbf{Avg Final}}   \\
 {} & \textbf{$\geq 0\%$} & \textbf{$>0.1\%$} & \textbf{$>1\%$} &
\textbf{Gain} & \textbf{Gain} \\ \midrule
Direct Optimisation  & 95\% & 89\%  & 46\% & 23.9\%  & 1.87\% \\ 
\rowcolor{Gray}Random Forest \cite{SPIERingis} (encoding features) & 67\% & 28\%   & 2\% & 8.5\%  & 0.19\% \\ \midrule 
 Random Forest (encoding new) & 68\% & 41\%   & 11\% & 12\%  & 0.23\% \\ 
\rowcolor{Gray} Random Forest (encoding + VGG features) & 69\% & 65\%   & 28\% & 18.3\%  & 0.82\% \\ \midrule 
\textbf{$k=0.7$ } & 58\% & 55\% & 25\%  & 9.5\% & 0.55\%  \\ 
\rowcolor{Gray}\textbf{$k=0.8$ } & 62\% & 56\% & 23\%  & 16.9\% & 0.58\%  \\ 
\textbf{$k=0.782$ } & 62\% & 56\% & 23\%  & 23.5\% & 0.63\%  \\ 
\midrule
\rowcolor{Gray}{DNN (encoding + VGG features)} & 89\% & 84\% & 39\%  & 14.9\% & 1.11\%  \\
\bottomrule
\end{tabular}
\end{table}

\subsection{Model \& Training}


\begin{table}[t]
    \caption{Details of the Dense Layer sizes ($n$), dropout (0.1) and batch-normalisation (BN). All layers use the Gaussian Error Linear Unit (GELU) activation function.}
    \setlength{\tabcolsep}{2pt}
    \centering 
    \begin{tabular}{lccccccccccc}\toprule
    layer id & 1 & 2 & 3& 4 & 5 & 6 & 7 & 8 & 9 & 10 & 11 \\ \midrule
 \rowcolor{Gray}   $n$ & 1000 & 800 & 600 & 200 & 100 & 64 & 32 & 16 & 8 & 4 & 1 \\
    BN & \checkmark & \checkmark &  &  & &  & & & & & \checkmark \\
\rowcolor{Gray}    Dropout & \checkmark & \checkmark & \checkmark &  & &  & &\checkmark & & &  \\\bottomrule
    \end{tabular}
    \label{tab:architecture}
\end{table}

Our model consists in a sequence of fully connected layers with GELU activation functions and Normalisation Layers. A full breakdown of the layers can be found in table \ref{tab:architecture}. The corpus is split 70\%/30\% between training and testing. The model was trained for 5000 epochs, using the SGD optimiser, a learning rate of 0.001 and a momentum of 0.9. The loss function is the mean absolute error between the predicted and directly optimised $k$.

\section{Results}

\begin{figure}
    \centering
    \includegraphics[height=0.7 \columnwidth]{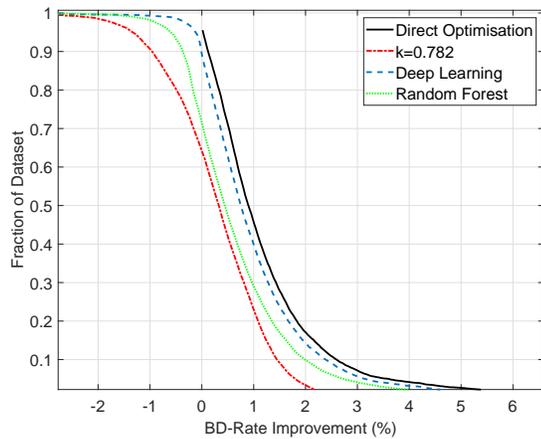}
    
    \caption{CDF plots of the BD-Rate improvement across our corpus for Random Forest \cite{SPIERingis} and our proposed DNN method. We see that both per-clip approaches outperform an overall adjustment of the Lagrangian multiplier ($k=0.782$) and that our network reaches near optimal performance.}
    \label{SmartResults}
\end{figure}

Both our proposed deep learning model, and the Random Forest system outlined in \cite{SPIERingis} (with the updated feature set) yield BD-Rate improvements comparable to the direct optimisation. Figure \ref{SmartResults} shows the portion of our test corpus which have a BD-Rate improvement. In these graphs, as well as the results presented in Table \ref{MLSummary}, we see that roughly 95\% of the clips had some BD-Rate improvement, with 46\% of the corpus having a 1\% or better BD-Rate improvement from direct optimisation. Our machine learning systems (Deep Learning and Random Forest from \cite{SPIERingis}) give positive results with 90\% and 70\% of the corpus getting a BD-Rate improvement but only 39\% and 28\% of the corpus getting a 1\% or better BD-Rate improvement. We can see that while all three random forest implementations had a similar portion of the corpus show improvement, the amount of improvement was best using both the expanded encoding features along with the VGG features. The Random Forest and deep learning systems have a very similar performance on clips which yield 5\% or better BD-Rate improvements. Both systems had roughly 1\% of the corpus giving gains of 5\% or more, but the Direct Optimiser was able to achieve 5\% or better gains in approximately 5\% of the clips. We believe that this is due to less training data available in the "high improvement" range. 

We also see that both machine learning systems drastically outperform the best global Lagrangian multiplier adjustment ($k=0.782$), which confirms the benefits of using a per clip approach.

Table \ref{MLSummary} also reports the {\em average positive BD-Rate improvement} as our {\em Average Final BD-Rate Gain}.
This is the gain measured when our system is used as a post-process or {\em optimised re-run} of the encoder given an initial encode with default settings. The best BD-rate between the initial encode and the re-run defines the final BD-Rate Gain. This is the most likely use of systems of this kind, and can even be considered as a two-pass system. This  allows us to take advantage of any improvements in BD-Rate from Lagrangian multiplier optimisation and default to the first default encode (unmodified Lagrangian Multiplier) in the case of worse performance. In this situation, we can see that our deep learning approach almost achieves what the direct optimiser does, at a fraction of the computational cost. The direct optimiser would have required on average sixty video encodes (12 RD-Curves) whereas our Deep Learning approach only required ten (2 RD-Curves) to get near comparable improvements.


\section{Conclusion}

In this paper, we have presented further data to support a per clip approach for the Lagrangian Multiplier within the HEVC codec. We build upon our direct optimisation results, by implementing a deep learning model capable of  comparable results in a fraction ($1/6$) of the computational cost. 90\% of our corpus shows some BD-Rate improvement with deep learning (compared to 95\% of the corpus with direct optimisation). We also see that we can get approximately 1\% BD-Rate savings across our entire corpus if we use the deep learning system as a supplement to the existing codec.  In summary, our deep learning approach nears the direct optimiser results while using only $1/6^{th}$ of the computational cost.
Future work will include investigating the deployment of this system for quality/distortion improvements at certain bitrates instead of BD-Rate gain as well as expansion to newer codecs like VVC.

\bibliographystyle{./bibliography/IEEEtran}
\bibliography{conference_041818}

\end{document}